\documentstyle[preprint,aps,epsf]{revtex}

\begin{document}
\preprint{
\font\fortssbx=cmssbx10 scaled \magstep2
\hbox to \hsize{
\hbox{\fortssbx University of Wisconsin - Madison}
\hfill$\vcenter{\hbox{\bf MADPH-97-988}
                \hbox{March 1997}
                \hbox{Revised July 1997}}$}
}

\title{\vspace{.5in}
Rigorous pion-pion scattering lengths from\\
threshold $\pi N\to \pi\pi N$ data}
\author{M. G. Olsson}
\address{Physics Department, University of Wisconsin, Madison, WI 53706}

\maketitle

\begin{abstract}
A new evaluation of the universal $\pi\pi$ scattering length relation is used to extract the $\pi\pi$ $s$-wave scattering lengths from threshold pion production data. Previous work has shown that the chiral perturbation series relating threshold pion production to $\pi\pi$ scattering lengths  appears to converge well only for the isospin-2 case, giving $a_2 = -0.031\pm 0.007 m_\pi^{-1}$. A model-independent and data-insensitive universal curve then implies $a_0 = 0.235\pm 0.03 m_\pi^{-1}$ for the isospin-0 scattering length.
\end{abstract}

\thispagestyle{empty}
\newpage

Pion-pion scattering is the simplest non-trivial hadron scattering process. It provides an ideal laboratory for the concepts  of chiral symmetry. The $\pi\pi$ scattering lengths, first predicted by Weinberg\cite{wein}, were improved by the development of chiral perturbation theory (ChPT) where higher order corrections could be computed\cite{gass}, recently to the two loop accuracy\cite{bijnens,knecht}. In the case of the $s$-wave $\pi\pi$ scattering lengths $a_I$, where isospin $I=0,2$, these higher-order corrections are reasonably small and the predictions are robust. It is the purpose of this paper to discuss how one can test these predictions using experimental data already at hand.

The principal present\footnote{The relatively rare decay $K\to\pi\pi\ell\nu$ can be used to extract the isospin $\delta_0-\delta_1$ phase shift difference below the $K$ mass\cite{ross}. Future experiments using this process, e.g.\ at DAPHNE, will accurately determine~$a_0$.} source of $\pi\pi$ scattering information is from the reaction $\pi N\to\pi\pi N$. For high-energy incident pions one can extrapolate to the kinematic region where single pion exchange is dominant\cite{charlie}. The most accurate experiments of this sort were done in the 70's by the CERN-Munich Collaboration\cite{hyams}. These phase shifts do not extend close enough to threshold to determine $a_0$ and $a_2$ directly, but they do provide a powerful constraint with the help of a dispersion relation. 

The close relation between the $\pi N\to\pi\pi N$ amplitude at pion production threshold and the $\pi\pi$ scattering lengths was point out long ago\cite{leaf}. In modern ChPT language this result is the lowest (tree level) order with no excited nucleon contributions.\footnote{The ChPT formalism requires $\xi=0$ in \cite{leaf}. This would yield the Weinberg $\pi\pi$ scattering lengths.} Recently the possibility of a more ambitious calculation of $\pi N\to\pi\pi N$ and its relation to the $\pi\pi$ scattering lengths has been discussed\cite{newbernard} and subsequently established\cite{bernard}. This result takes into account loop corrections as well as $\pi N$ resonant effects. In summary, it was found that the isospin-2 scattering length was related to the threshold $\pi N\to \pi\pi N$ amplitude in almost exactly the same way as the lowest order result\cite{leaf}. The ChPT series in this case appears to be very convergent and the loop and resonance corrections are small. On the other hand, resonance and higher order corrections make important changes in the relation of the isospin-0 scattering length and the threshold production amplitude. For $a_0$ one has little confidence that the $\pi N\to \pi\pi N$ threshold extrapolation will yield a believable result within the present scheme\cite{bernard}.

Several years ago a remarkable series of experiments were done at several laboratories measuring pion production very near to threshold\cite{kernel}. A global analysis was done\cite{bark} to extract the two isospin threshold amplitudes. The result has now been interpreted\cite{bernard} in terms of the isospin-2 scattering length as
\begin{equation}
a_2 = -0.031\pm 0.007 m_\pi^{-1} \,.  \label{eq:a_2}
\end{equation}

Analyticity constraints using $\pi\pi$ phase shifts in the energy region $0.5<\sqrt s<2$~GeV also do not fully determine the the $s$-wave scattering lengths. This is because threshold subtractions are needed to obtain convergence of the crossing even sum rules. It was noticed\cite{morgan,martin} though that, from either forward dispersion relations or from the Roy equations, $a_0$ and $a_2$ were constrained to lie on a {\it universal curve}. Thus when one scattering length is known the other is provided. The content of the universal curve is given in a concise way by the $t$-channel isospin-1 forward dispersion relation evaluated at threshold\cite{martin},
\begin{equation}
2a_0 - 5a_2 = {12\over\pi} \int_0^\infty {dq\over q(1+q^2)} {\rm Im}\, A_{I_t=1}(q) \,,  \label{eq:sumrule}
\end{equation}
where $q$ is the c.m.\ pion momentum in units of $m_{\pi^\pm}$. In terms of direct channel isospin amplitude $A_{I_s}$ we have\cite{martin}
\begin{equation}
A_{I_t=1} = {1\over3} A_{I_s=0} + {1\over2} A_{I_s=1} - {5\over6} A_{I_s=2}\,.
\end{equation}
The amplitude is normalized as
\begin{eqnarray}
A_{I} &=&  {\sqrt{1+q^2}\over q} \sum_{\ell=0}^\infty (2\ell+1) f_\ell(q) P_\ell(\cos\theta=1) \,,\\
f_\ell(q) &=& (\eta_\ell e^{2i\delta_\ell} - 1) / 2i \,,
\end{eqnarray}
and hence in the threshold limit
\begin{equation}
A(q) \to \delta_{\ell=0} / q \to a_{\ell=0} \,,
\end{equation}
so the invariant amplitude at threshold is the $s$-wave scattering length. The crossing odd sum rule (\ref{eq:sumrule}) converges without subtraction as Regge rho exchange should give Im$\,A_{I_t=1} \stackrel{q\to\infty}{\longrightarrow} cq$.

In earlier work\cite{chell} we have used subtracted dispersion relations together with the CERN-Munich phase shifts (for $\sqrt s > 0.5$~GeV) and assumed $s$-wave scattering lengths ($a_0=0.20 m_\pi^{-1}, a_2 = -0.032 m_\pi^{-1}$) to interpolate smoothly between threshold and $\sqrt s = 0.5$~GeV. We now explictly remove the scattering length dependence by assuming that near threshold the $s$-wave phase shifts can be represented by
\begin{equation}
\tan\delta_{\ell=0} = {aq\over 1+q^2}
\end{equation}
and hence
\begin{equation}
{\rm Im}\, A_{\ell=0} ={qa^2\sqrt{1+q^2} \over  \left(1+q^2\right)^2 + a^2q^2} \,.  \label{foo}
\end{equation}
We subtract this from the interpolated data and then add a similar expression with an arbitrary scattering length\footnote{If the actual scattering lengths are nearly the same as assumed in the interpolated data\cite{chell} (as is the case), this introduces little or no error.} which we insert into the sum rule (\ref{eq:sumrule}) and explicitly integrate. The sum rule (\ref{eq:sumrule}) then takes the form\footnote{In the integration of (\ref{foo}),  terms of order $a^4$ have been dropped}
\begin{equation}
2\left( a_0 - {4\over3\pi} a_0^2 \right) - 5\left(a_2 - {4\over3\pi}a_2^2\right) = L_0 \,, \label{eq:sumrule2}
\end{equation}
\begin{equation}
L_0 = {12\over\pi} \int_0^\infty {dq\over q(1+q^2)} \; {\rm Im} \, A_{I_t=1}^{(0)} (q) \,,  \label{eq:L_0}
\end{equation}
where $A^{(0)}$ is the CERN-Munich data extrapolated to zero $s$-wave scattering lengths. The scattering length dependence now appears explicitly on the left-hand side of (\ref{eq:sumrule2}). The integrand of the sum rule (\ref{eq:L_0}) from the $\ell=0$ and $\ell=1$ partial waves is depicted in Fig.~1. The dominance of the $\rho(770)$ is evident. Evalution of (\ref{eq:L_0}) is now straightforward and the result is
\begin{equation}
L_0 = 0.58 \pm 0.015 \, m_\pi^{-1} \,.   \label{eq:result}
\end{equation}
A conservative error has been estimated by assuming that the quoted errors on the phase shifts (about $1^\circ$) are entirely systematic. If it were statistical the error of the integral would be very small. The largest contribution to (\ref{eq:result}) comes from the rho state. Increasing $\delta_1$ by one degree raises $L_0$ by 0.007. The higher resonances which couple to $\pi\pi$ scattering (and their contributions to $L_0$) are\cite{pdg}: $f_2(1270) = 0.051$, $\rho_3(1690) = 0.014$, and $f_4(2050) = 0.006$.

In Fig.~2 we show (\ref{eq:sumrule2}) with $L_0$ as in (\ref{eq:result}). This is virtually identical with the old (pre CERN-Munich) universal curve\cite{morgan,martin}. The corridor represents the assigned error in $L_0$ given in (\ref{eq:result}). The vicinity where (\ref{eq:a_2}) $a_2=-0.031\pm 0.007$ would lie on the universal curve is shown in Fig.~3. Taking into account both the errors in $a_2$ and $L_0$ we conclude that the isospin-0 scattering length must be
\begin{equation}
a_0 = 0.235\pm 0.03 m_\pi^{-1} \,.  \label{eq:a_0}
\end{equation}
It is evident that reducing the error in $a_2$ by a more thorough ChPT analysis would pay considerable dividends.

\section*{Summary and Conclusions}

A ChPT analysis\cite{bernard} has shown that the isospin-2 $\pi\pi$ scattering length $a_2$ can be extracted from threshold $\pi^+p\to\pi^+\pi^+ n$ data. A new evaluation of the universal curve then provides a rigorous value (12) for the isospin-0 scattering length $a_0$.

To compare our result with ChPT predictions for the scattering lengths we begin with the lowest order result of Weinberg\cite{wein}
\begin{eqnarray}
a_0 &=& {7\over4}L = 0.158 m_\pi^{-1} \,, \label{wein-a}\\
a_2 &=& -{1\over2}L = -0.045 m_\pi^{-1} \,, \label{wein-b}
\end{eqnarray}
where $L=1/8\pi F_\pi^2 = 0.090$ with $F_\pi=93$~MeV. We note that both are several multiples of the assigned errors from the experimental values for $a_0$ in (\ref{eq:a_0}) and $a_2$ in (\ref{eq:a_2}). We also note that the Weinberg scattering length does not lie on the universal curve as seen in Fig.~3. The one loop corrections\cite{gass,bijnens} to the scattering lengths involve four renormalization constants relating to $m_\pi$, $F_\pi$, and to higher resonances. Actually only two independent combinations of these constants enter the expression for $a_0$ and $a_2$. Using the values of these constants found from other experimental evidence\cite{bijnens} one obtains
\begin{eqnarray}
a_0 &=& \phantom+0.200 m_\pi^{-1} \,, \label{last-a}\\
a_2 &=& -0.043 m_\pi^{-1} \,. \label{last-b}
\end{eqnarray}
The above one loop ChPT prediction is depicted in Fig.~3. We observe that the prediction falls within the universal curve and that it is not inconsistent with (but just outside) the point obtained from threshold pion production and the universal curve.

The $\pi\pi$ amplitude has been recently evaluted to the two loop level\cite{gass,bijnens}. The result\cite{gass} for $a_0$ and $a_2$ is also depicted in Fig.~3. It should be noted that the contribution of this level is relatively small. It should also be pointed out that since the pion production calculation\cite{bernard} has been calculated to one loop accuracy the $\pi\pi$ scattering amplitude can only be tested to this order by our result. 

The requirement that the $(a_0,a_2)$ point lie on the universal curve is general. At the one loop level we observe that\cite{gass}
\begin{equation}
2a_0 - 5a_2 = (0.55 + 0.0155\bar\ell_4) m_\pi^{-1} \,. \label{oneloop-a}
\end{equation}
At $a_0 = 0.2 m_\pi^{-1}$ the universal curve $2a_0-5a_2 = (0.61\pm 0.015)m_\pi^{-1}$ which gives
\begin{equation}
\bar\ell_4 = 3.9\pm 1.0 \,.  \label{oneloop-b}
\end{equation}
This is in excellent agreement with the value $\bar\ell_4 = 4.6\pm 1.2$ obtained by considerations of the $\pi\pi$ scalar radius\cite{gass-new}.

One might ask, why doesn't the two-loop ($a_0, a_2$) point lie within the universal curve corridor? The answer is that there are additional subtraction constants which enter even the combination $2a_0-5a_2$. This is illustrated in a more complete presentation\cite{bij2} by the authors of Ref.~\cite{bijnens}, who determine two sets of these constants. Set~I falls in the universal corridor while set~II is similar to their previous set\cite{bijnens} illustrated in Fig.~3. Even at the one-loop level we observed in (\ref{oneloop-a}) and (\ref{oneloop-b}) that the parameter $\bar\ell_4$ must be chosen rather carefully for ($a_0,a_2$) to lie in the universal corridor. 

It will thus be of considerable interest to improve the result (1) for $a_2$ since about half the quoted error\cite{bernard} arises from the theoretical analysis. The present result for $a_0$ would also be automatically improved.

\section*{Acknowledgments}
I would like to think Sini\v sa Veseli for helpful comments.
This research was supported in part by the U.S.~Department of Energy under Grant No.~DE-FG02-95ER40896 and in part by the University of Wisconsin Research Committee with funds granted by the Wisconsin Alumni Research Foundation.

\newpage
\section*{Figure Captions}

\begin{itemize}
\item[Fig.~1:] $L_0$ integrand as a function of c.m.\ pion momentum. Shown are $\ell=0$ and 1 contributions only. The integrand is dominated by the rho(770) state.

\item[Fig.~2:] The universal curve for pion-pion $s$-wave scattering lengths as computed using (9) with $L_0$ given by (\ref{eq:L_0}). The corridor represents the error assignment on $L_0$.

\item[Fig.~3:] Portion of the universal curve near the physical values. The experimental point for $a_2$ is from (\ref{eq:a_2}). The horizontal error on this point indicates the uncertainty in $a_0$ implied by the universal curve. The tree (Weinberg) values (\ref{wein-a},\ref{wein-b}) are indicated and the one and two loop chiral perturbation theory predictions  are noted. \end{itemize}

\newpage
\centering
\vspace*{1.5in}

\hspace{0in}\epsffile{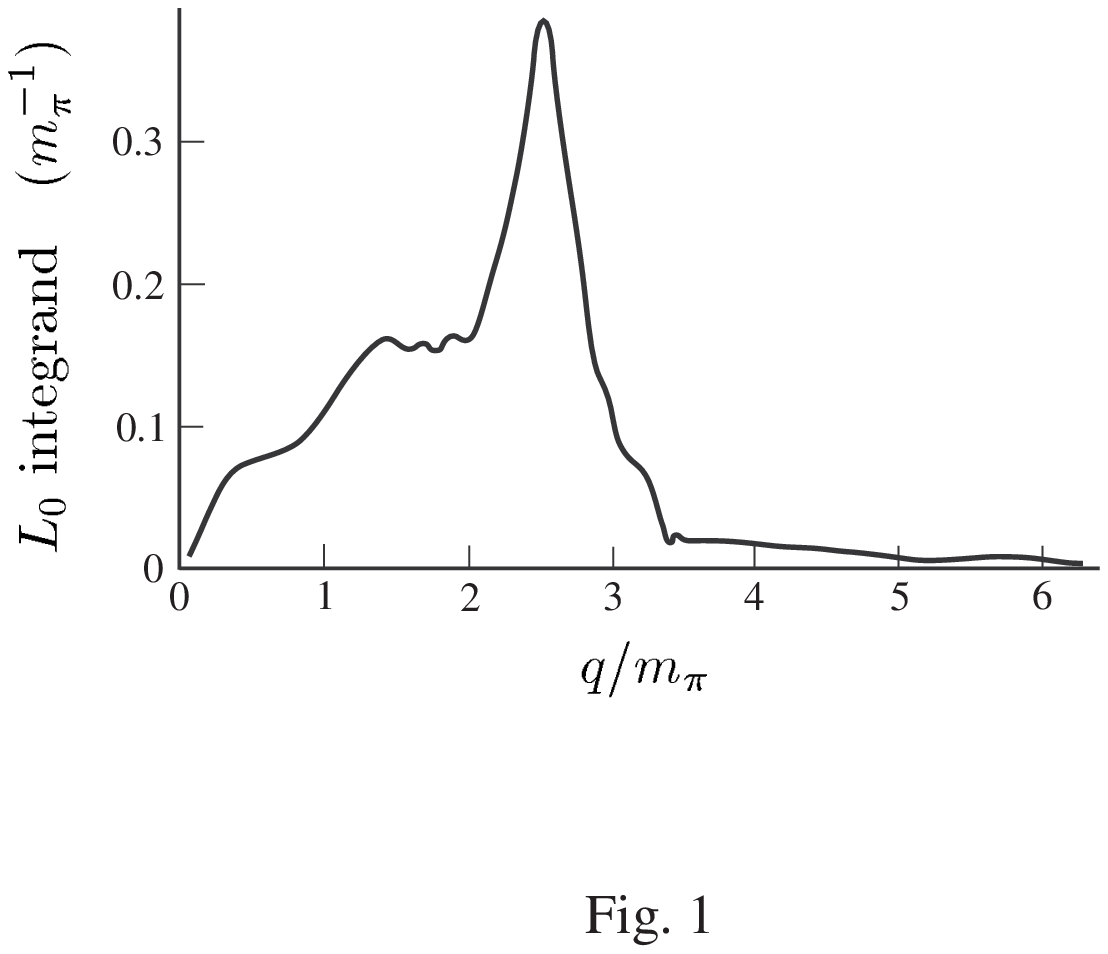}

\newpage
\vspace*{1.5in}

\hspace{0in}\epsffile{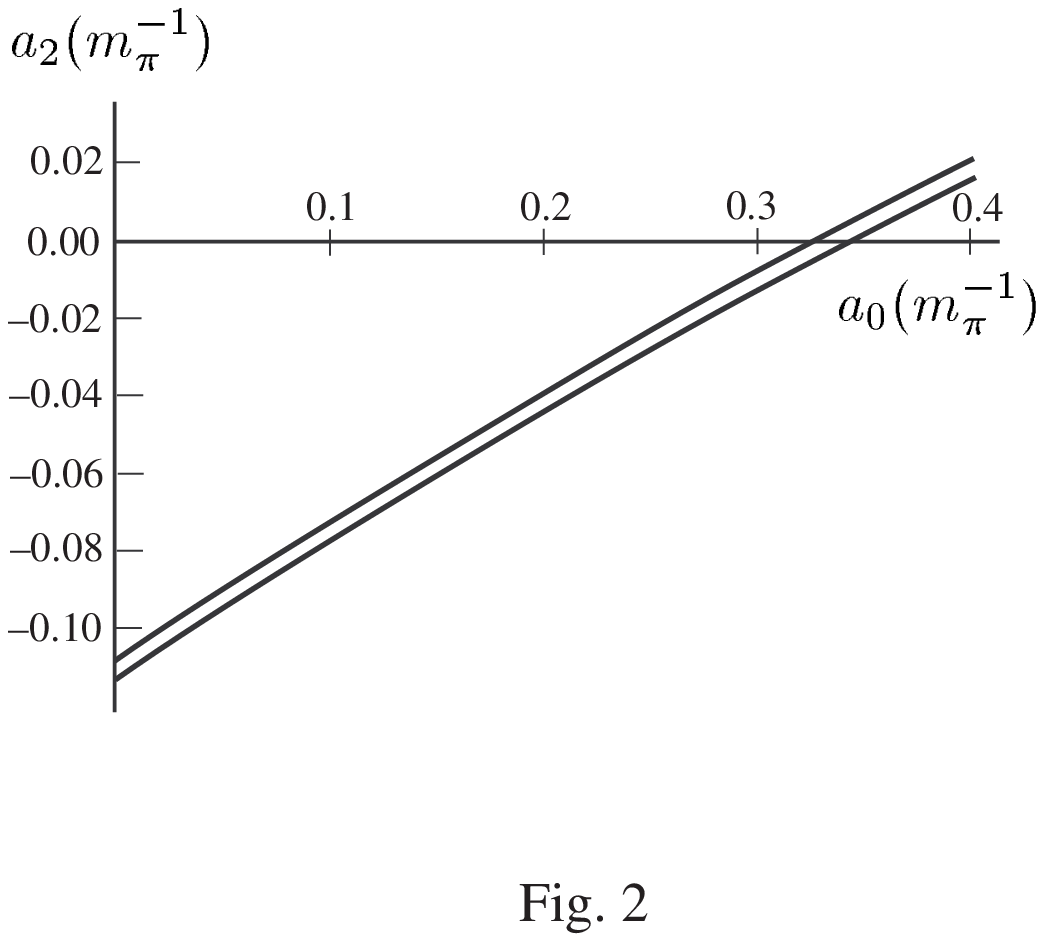}

\newpage
\vspace*{1.5in}

\hspace{0in}\epsffile{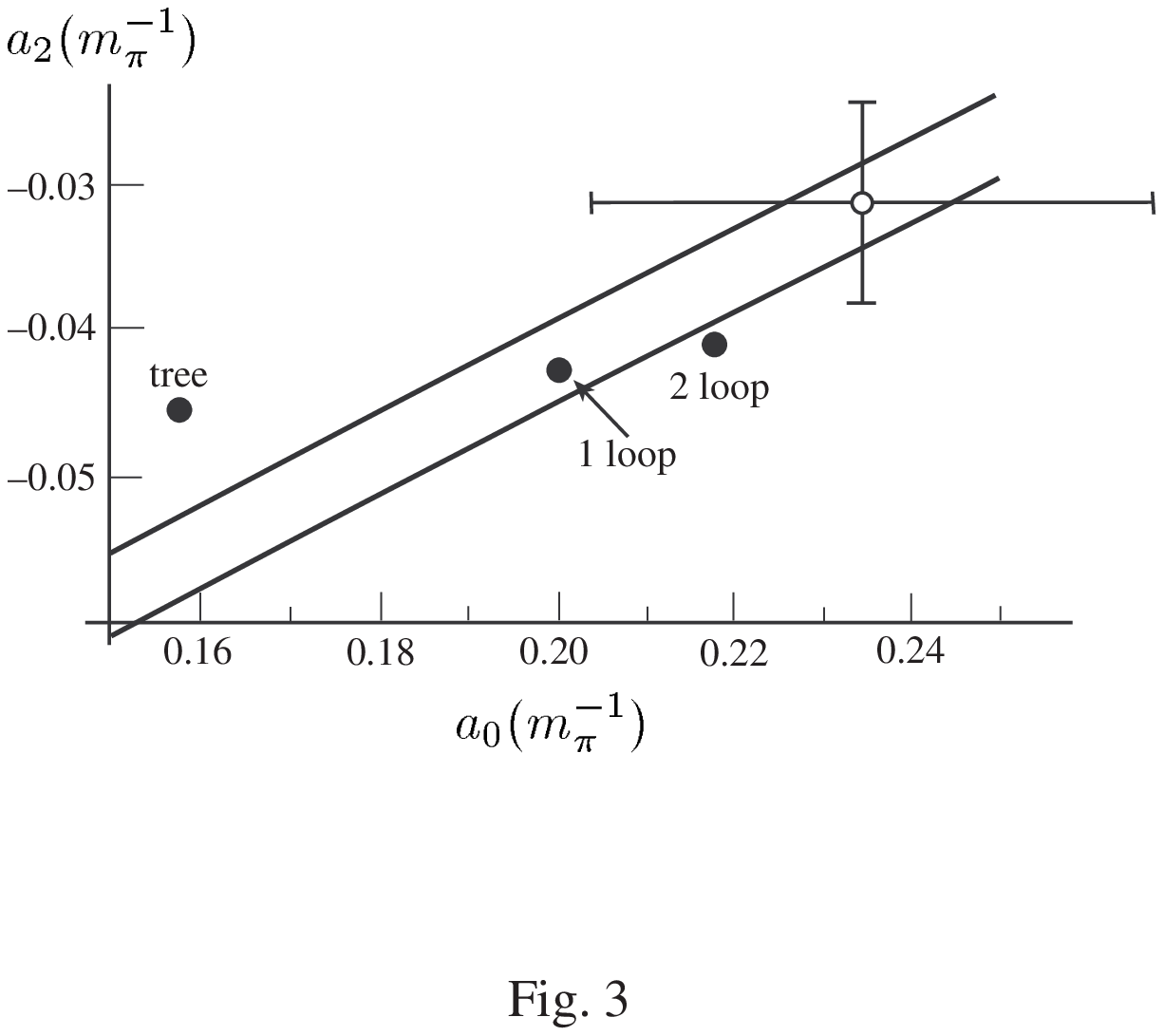}

\end{document}